# Spin Hall effect from hybridized 3*d*-4*p* orbitals

Yong-Chang Lau[1,2†**], Hwachol Lee[2†], Kohji Nakamura[3] and Masamitsu Hayashi[1,2**]

[1]*Department of Physics, The University of Tokyo, Bunkyo, Tokyo 113-0033, Japan*

[2]*National Institute for Materials Science, Tsukuba 305-0047, Japan*

[3]*Department of Physics Engineering, Mie University, Tsu, Mie 514-8507, Japan*

[†]These authors contributed equally to this work

Electrical manipulation of magnetization by spin-orbit torque (SOT)[1-6] has shown promise for realizing reliable magnetic memories and oscillators. To date, the generation of transverse spin current and SOT, whether it is of spin Hall effect (SHE)[2,7], Rashba-Edelstein effect[8] or spin-momentum locking origin[3,9], relies primarily on materials or heterostructures containing 5*d* or 6*p* heavy elements with strong spin-orbit coupling. Here we show that a paramagnetic CoGa compound possesses large enough spin Hall angle to allow robust SOT switching of perpendicularly-magnetized ferrimagnetic MnGa films in CoGa/MnGa/Oxide heterostructures. The spin Hall efficiency estimated via spin Hall magnetoresistance and harmonic Hall measurements is +0.05±0.01, which is surprisingly large for a system that does not contain any heavy metal element. First-principles calculations corroborate our experimental observations and suggest that the hybridized Co 3*d* – Ga 4*p* orbitals along R-X in the Brillouin zone is responsible for the intrinsic SHE. Our results suggest that efficient spin current generation can be realized in intermetallic by alloying a transition metal with a *p*-orbital element and by Fermi level tuning.

[*] yongchang.lau@qspin.phys.s.u-tokyo.ac.jp

[**] hayashi@phys.s.u-tokyo.ac.jp

Generation of spin current via the spin Hall effect is considered one of the most viable approach toward realizing SOT-based devices. Recent experiments have demonstrated efficiency[2,10-13] of spin current generation that exceeds ~10%, largely in a class of materials that contains a heavy metal element. Although the underlying mechanism of the spin Hall effect in such materials is under debate, it is clear that the large spin orbit coupling of the heavy metals plays a dominant role in the spin current generation process[14,15]. However, the necessity to include elements with large spin orbit coupling limits the choice of material to explore systems with possibly large spin current generation efficiency.

Here, we show that $\beta$-CoGa compound, which is non-magnetic at room temperature, generates spin current that allows electrical control of the magnetization of a neighboring magnetic layer. The spin Hall efficiency is determined to be +0.05±0.01, surprisingly large given that the layer contains no heavy metals with strong spin orbit coupling. First principles calculations suggest that the spin Hall conductivity of CoGa is 140 (ħ/e) $\Omega^{-1}$cm$^{-1}$, which is in good agreement with the experimentally estimated value of 143±30 (ħ/e) $\Omega^{-1}$cm$^{-1}$. The calculations demonstrate that the spin Hall conductivity in CoGa largely originates from a band which has a character of mixed Co $3d$ − Ga $4p$ orbitals. Using CoGa as a seed layer, we demonstrate magnetization switching of Mn-based ferrimagnetic tetragonal Heusler compound[16] (MnGa), a material system that can be applied to ultrafast spintronic applications owing to its high perpendicular magnetic anisotropy (PMA)[17,18], low Gilbert damping[19], high spin polarization[20] and large tunneling magnetoresistance (TMR)[21]. These results are not only of high technological potential but also provides a different approach toward generation of large spin current for current and voltage controlled magnetization in thin film heterostructures.

Figure 1(a) defines the coordinate system and illustrates our typical stacks, which consist of MgO(001) substrate/Co$_{53}$Ga$_{47}$(CoGa)($t$)/FM/(Mg$_2$Al)O$_y$(MAO)(2)/Ta(1) (in nm). FM denotes either a 2 nm-thick ferrimagnetic Mn$_{1.44}$Ga (MnGa) with PMA or a 1 nm-thick in-plane magnetized ferromagnetic Co$_{20}$Fe$_{60}$B$_{20}$ (CoFeB) layer. Equiatomic $\beta$-Co$_{50}$Ga$_{50}$ is known to crystallize in CsCl-type $B2$ structure (Space group 221), as illustrated in Figure 1(b), with Co and Ga occupying equivalent l$a$ (0, 0, 0) and l$b$ (0.5, 0.5, 0.5) sublattices, respectively. Off-stoichiometric Co(Ga)-rich compounds may retain the $B2$ structure by forming Co antisites (vacancies) in lieu of Ga (Co)[22]. Symmetrical $\theta$-$2\theta$ X-ray diffraction spectrum of a CoGa(5)/MnGa(10) heterostructure is plotted in Figure 1(c). The appearance of CoGa(001) superlattice peak confirms the $B2$ ordering. Detailed peak intensity analysis suggests simultaneous presence of ~10 % Co antisites and vacancies (Supplementary S1). The pole figure in Figure 1(d) reveals the in-plane epitaxial relationship of the structure: MgO[110]//CoGa[100]//MnGa[100], with CoGa and MnGa making a 45$^\circ$ in-plane rotation with respect to the MgO lattice. The SQUID-VSM magnetometry of a CoGa(5)/MnGa(2) heterostructure is shown in Figure 1(e). We find saturation magnetization $M_s$ of ~400 emu/cm$^3$ and anisotropy field $H_k$ of ~20 kOe, corresponding to an effective uniaxial anisotropy $K_{\text{eff}} \approx M_s H_k/2 \sim 4$ Merg/cm$^3$.

The Hall resistance $R_{xy}$ as a function of the out of plane field ($H_z$) for a Hall bar device patterned from the CoGa(5)/MnGa(2) heterostructure is shown in Figure 1(f). The square $R_{xy}$-$H_z$ loop confirms the strong PMA of the magnetic layer with a coercive field $H_c \sim 1.2$ kOe. SOT-induced switching of MnGa is demonstrated by sweeping the current $I$ along $x$ while monitoring the $z$-component of the magnetization via $R_{xy}$. $R_{xy}$-$I$ loops, measured with a bias field $H_x = \pm 1$ kOe, are plotted in Figure 1(g) and 1(h). Magnetization along $-z$ ($+z$) is preferred when the current flow and $H_x$ are parallel (anti-parallel), which suggests that the spin Hall angle of CoGa is positive (same sign with that of Pt[12]). Thus a charge current flowing along $+x$ generates a spin accumulation ($\boldsymbol{\sigma}$) polarized along $-y$ at the top interface of CoGa via the spin Hall effect (SHE). The switching current $I_{sw} \sim 11$ mA corresponds to a current density $J_{sw} \sim 1.6 \times 10^7$ A/cm$^2$.

The efficiency $\xi_{SL}$ of the Slonczewski-like spin Hall torque acting on the adjacent magnetization $\boldsymbol{m}$ is linked to the intrinsic spin Hall angle $\theta$ by: $\xi_{SL} = T \cdot \theta$ where typically $T (\leq 1)$ describes the interfacial spin transparency[23]. From hereon we focus on quantifying $\xi_{SL}$ and its field-like counterpart $\xi_{FL}$ of CoGa/MnGa and CoGa/CoFeB heterostructures to obtain the lower bounds of $\theta_{CoGa}$. First we discuss $\xi_{SL}$ obtained from magnetoresistance measurements. The longitudinal resistance ($R_{xx}$) of a heavy metal/ferromagnetic metal bilayer can be modulated by the transmission and reflection of SHE-induced spins at the interface due to the collective action of the SHE and the inverse SHE. Owing to this effect, now commonly referred to as the spin Hall magnetoresistance (SMR), $R_{xx}$ shows a distinct difference when $\boldsymbol{m}$ is directed parallel or perpendicular to $\boldsymbol{\sigma}$.[24-26]

Figure 2(a) and 2(b) show the magnetoresistance (MR) ratio, $\Delta R_{xx}(H_i)/R^i_{xx,0}; i = \boldsymbol{x}, \boldsymbol{y}, \boldsymbol{z}$, of CoGa(5)/Mn$_{1.44}$Ga(2) and CoGa(5)/CoFeB(1) heterostructures with magnetic field applied along three orthogonal axes: $\boldsymbol{x}, \boldsymbol{y},$ and $\boldsymbol{z}$. Here, $R^i_{xx,0} \equiv R_{xx}(H_i = 0)$ is the base resistance and $\Delta R_{xx}(H_i) \equiv R_{xx}(H_i) - R^i_{xx,0}$ is the MR due to an applied field along the $i$ direction. For both heterostructures, the MR is dominated by a background signal that does not depend on the magnetization direction of the MnGa or CoFeB layer. We find that the background MR mainly arises from the paramagnetic CoGa layer (see Supplementary S2). This contribution can be largely eliminated by plotting instead $\Delta R^{x-z}_{xx}/R^z_{xx,0} \equiv (R_{xx}(H_x) - R_{xx}(H_z))/R^z_{xx,0}$ and $\Delta R^{y-z}_{xx}/$ $R_{xx,0}^z \equiv (R_{xx}(H_y) - R_{xx}(H_z))/R_{xx,0}^z$ , as shown in Figure 2(c) and 2(d) for the two heterostructures. The $\boldsymbol{m}$-dependent contribution of $R_{xx}^{x-z}$ and $R_{xx}^{y-z}$, which are commonly known as anisotropic magnetoresistance (AMR) and SMR, respectively, are extracted by extrapolating the field-dependent MR contributions to zero field. Upon including the MR contribution of MnGa, $\Delta MR_{MnGa}$ as detailed in the supplementary S3, we estimate the magnitude of the SMR ratio $\Delta SMR$, as illustrated in Figure 2(c) and 2(d). The CoGa thickness $t$ dependence of $\Delta SMR$ for the two heterostructures is plotted in Figure 2(e). Assuming a transparent interface, we fit the data with an expression derived from a drift-diffusion model[26,27]:

$$\Delta SMR = \xi_{CoGa/FM,SL}^2 \frac{\lambda}{t} \frac{\tanh(t/2\lambda)}{1+a} \left[ 1 - \frac{1}{\cosh(t/\lambda)} \right] \quad (1)$$

We use the spin diffusion length of CoGa $\lambda$ and $\xi_{SL}$ as the fitting parameters. $a \equiv \rho_{CoGa} t_{FM}/\rho_{FM} t$ describes the current shunting due to the presence of conducting magnetic layer where $\rho_{CoGa}$ is the resistivity of CoGa and $\rho_{FM}$ denotes the resistivity of the magnetic layer of thickness $t_{FM}$. We obtain $|\xi_{CoGa/MnGa,SL}| \sim |\xi_{CoGa/CoFeB,SL}| \sim 0.05$ and in average $\lambda \sim 1.8$ nm with the fitted curves shown in Figure 2(e).

We next quantify the Slonczewski-like ($\boldsymbol{H}_{SL} \parallel \boldsymbol{m} \times \boldsymbol{\sigma}$) and the field-like ($\boldsymbol{H}_{FL} \parallel -\boldsymbol{\sigma}$) spin-orbit effective fields in the two heterostructures using the harmonic Hall techniques[28-30]. Under a low frequency sinusoidal excitation current density of amplitude $J_o$, the current-dependent effective fields (*i.e.* the SOT) modulate the direction of $\boldsymbol{m}$, thereby producing an out-of-phase second harmonic Hall resistance, $R_{2\omega}$. To estimate the effective fields, we use field scans that depend on the magnetization easy axis of the heterostructures.

For the perpendicularly magnetized CoGa($t$)/MnGa(2) heterostructures, the change of the in-phase first harmonic (fundamental) Hall resistance $R_\omega$ and $R_{2\omega}$, in response to a moderate $H_x$ or $H_y$, are recorded (Figure 3(a)). Typical $R_{2\omega}$ against $H_x$ and $H_y$ for a sample with $t = 5$ nm are plotted in Figure 3(b) and 3(c), respectively. We define $B_x \equiv \frac{\partial R_{2\omega}(H_x)}{\partial H_x} / \frac{\partial^2 R_\omega(H_x)}{\partial H_x^2}$, $B_y \equiv \frac{\partial R_{2\omega}(H_y)}{\partial H_y} / \frac{\partial^2 R_\omega(H_y)}{\partial H_y^2}$ to obtain the spin-orbit effective fields as the following[31]:

$$H_{SL} = -2\frac{B_x \pm 2\epsilon B_y}{1-4\epsilon^2} \; ; \; H_{FL+Oe} = -2\frac{B_y \pm 2\epsilon B_x}{1-4\epsilon^2} \quad (2)$$

The $\pm$ sign corresponds to $\boldsymbol{m}$ pointing along $\pm z$. A positive $H_{SL}$ ($H_{FL+Oe}$) represents $\boldsymbol{H}_{SL}$ ($\boldsymbol{H}_{FL+Oe}$) pointing along $+x$ ($+y$). $\epsilon (\equiv \Delta R_{PHE}/\Delta R_{xy})$ denotes the ratio of the magnitude of planar Hall effect (PHE) $\Delta R_{PHE}$ and that of the anomalous Hall effect (AHE) $\Delta R_{xy}$. From high field measurement, we estimate $\varepsilon \sim 0.11$. Here $H_{FL+Oe}$ includes Oerstend field contribution. We evaluate the Oersted field ($H_{Oe}$) arising from the CoGa layer and acting on $\boldsymbol{m}$: the calculations return $H_{Oe}/J_0 \approx$ -0.31 Oe/$10^6$Acm$^{-2}$ for $t = 5$ nm. We subtract $H_{Oe}$ from $H_{FL+Oe}$ to obtain $H_{FL}$. Slopes of linear regressions on $H_{SL}$ and $H_{FL}$ against $J_0$ are used to evaluate the efficiencies $\xi_{SL} = \frac{2e}{\hbar} \frac{H_{SL}M_s t_{FM}}{J_0}$ and $\xi_{FL} = \frac{2e}{\hbar} \frac{H_{FL}M_s t_{FM}}{J_0}$. For $t = 5$ nm we find $\xi_{CoGa(5)/MnGa,SL} \approx +0.034 \pm 0.020$ and $\xi_{CoGa(5)/MnGa,FL} \approx +0.27 \pm 0.10$. $\xi_{CoGa(5)/MnGa,SL}$ is smaller than that estimated using the SMR method; however this is due to the fact that a CoGa thickness of 5 nm is not large enough, compared to its spin diffusion length ($\lambda \sim 1.8$ nm), to observe saturation of the effective field (see Figure 3(f) and the related description below).

For in-plane magnetized CoGa($t$)/CoFeB(1) bilayers, the in-phase first harmonic Hall resistance $R_\omega$ and $R_{2\omega}$ are measured as a function of the angle $\varphi$ between the current flow (along $\boldsymbol{x}$) and the magnetic field applied within the film ($xy$) plane (Figure 3(d)). Spin-orbit effective fields are estimated as described by Avci *et al.*[30] (Supplementary S5). The CoGa thickness ($t$) dependence of $H_{SL}/J_o$ and $H_{FL}/J_o$ are plotted in Figure 3(e). Taking into account the CoGa thickness dependence of CoFeB $M_s$, ranging from 1100 to 1700 emu/cm³, the efficiency of the Slonczewski-like ($\xi_{CoGa/CoFeB,SL}$) and field-like ($\xi_{CoGa/CoFeB,FL}$) torques are plotted in Figure 3(f). We fit the thickness dependence with the expression[32]:

$$\xi_{CoGa/CoFeB,SL(FL)}(t) = \xi_{CoGa/CoFeB,SL(FL)}[1 - \text{sech}(t/\lambda)] \quad (3)$$

which gives $\xi_{CoGa/CoFeB,SL} = +0.053$ and $\lambda = 2.0$ nm, in satisfactory agreement with the SMR method. We find $\xi_{CoGa/CoFeB,FL}$ similar in magnitude with $\xi_{CoGa/CoFeB,SL}$. Interestingly, $\xi_{SL}$ is similar in magnitude for both bilayers CoGa/MnGa and CoGa/CoFeB, whereas $\xi_{FL}$ is significantly larger for the former. These results suggest that SOT and $\xi$ sensitively depend on the interfacial band matching, *i.e.* the interface transparency. We attribute the exceptionally large $\xi_{CoGa/MnGa,FL}$ to the significant imaginary part of the interfacial spin mixing conductance, which may include contribution from Rashba-like effects[33,34].

It is commonly believed that the spin-orbit coupling, scaling with $Z^4$ ($Z$ = atomic number), is an essential ingredient for obtaining large SOT. However, our observations on CoGa compound exhibiting appreciable SHE at room temperature challenge this archetypal assumption. The high $\rho_{CoGa}$ with an estimated mean free path $l$ comparable to twice of the lattice spacing $d \sim 2.8$ Å (Supplementary information S6) in general excludes interpretation based on extrinsic spin-dependent skew scattering. In order to gain insights into the mechanism of spin current generation in CoGa, we perform first-principles calculations based on density-functional theory to study the *intrinsic* spin Hall conductivity (SHC) of stoichiometric $\beta$-CoGa. We find that the hybridized Co 3$d$ − Ga 4$p$ orbitals along R-M in the Brillouin zone around the Fermi level (Figure 4(c)) is the primary source of the spin Berry curvature (Figure 4(a)). Upon integrating the latter over occupied states in all k-space, we obtain a calculated SHC of +140 (ℏ/e) $\Omega^{-1}$ cm$^{-1}$. This is in excellent agreement with the experimental SHC of +143±30 (ℏ/e) $\Omega^{-1}$ cm$^{-1}$, obtained by taking $\rho_{CoGa}$ = 175 $\mu\Omega$ cm and $\xi_{CoGa}$ ~ +0.05±0.01. The analysis of the SHC within a rigid band model, as shown in Figure 4(d), moreover indicates that the SHC increases when the energy is going from above the Fermi level to lower energy, *e.g.*, the SHC increases by ~50% if one could lower the Fermi level by a few tenths of eV.

Recently, strong intrinsic AHE and SHE has also been predicted in another heavy-element-free compound Mn$_3$Ge in a frustrated triangular lattice[35-37]. Here, using CoGa as an example, we demonstrate the critical role of $d - p$ orbital hybridization and Fermi level tuning in obtaining large Berry curvature and consequently large intrinsic SHE in intermetallics, without necessarily invoking heavy elements with strong spin-orbit coupling. Isostructural compounds containing $d$ and $p$ elements represent a vast material class (including but not limited to Heusler compounds[16]) that offers enormous possibilities. Exotic materials with non-trivial topology at the Fermi level can therefore be designed and realized through appropriate selection of the main chemical constituents and dopants.

The appreciable spin Hall effect in CoGa together with its unique capability for stabilizing an ultrathin MnGa with strong PMA in CoGa/MnGa/Oxide heterostructures allows realization of magnetic switches and oscillators with significantly higher thermal stability and operating speed than the common Ta/CoFeB/MgO structure. The anisotropy and the resonance frequency of MnGa can be easily tuned by varying the Mn:Ga ratio or substituting Ge for Ga, thus providing an attractive solution for bridging the "TeraHertz gap". Moreover, we expect further enhancement of the competitiveness of our structure by combining SOT-induced switching with voltage-controlled magnetic anisotropy at the MnGa/Oxide interface.

## Acknowledgements

We would like to thank S. Mitani for instrumental help, G. Qu for preliminary first-principles calculations and fruitful discussions. This work was partly supported by JSPS Grant-in-Aids for Specially Promoted Research (15H05702), MEXT R & D Next-Generation Information Technology, and Innovation and Spintronics Research Network of Japan. Y.-C.L. is a JSPS international research fellow.

## Author contributions

Y.-C.L. and H.L. contributed equally to this work. Y.-C.L. and H.L. conceived the experiment and planned the study with help of M.H. H.L. sputtered the films, optimized the structure and measured magnetic and structural properties of the unpatterned films. Y.-C.L. fabricated the devices, performed electrical measurements and analyzed the data with help of M.H. K.N. performed the first-principles calculations. Y.-C.L. and M.H. wrote the manuscript. All authors discussed the results and commented on the manuscript.

## Methods

*Film growth and device fabrication*

All the films are grown on MgO(001) substrates in a ultrahigh vacuum chamber by magnetron sputtering. CoGa layers are deposited at $200^{\circ}$C substrate temperature and postannealed at $400^{\circ}$C for obtaining $B2$ structure, unless otherwise stated. The remaining layers are grown at ambient temperature. The top Ta(1) layer is fully oxidized and is considered to be insulating. Hall bar devices with a nominal channel width of 10μm and a length of 25μm between the Hall probes for measuring longitudinal resistance are fabricated by standard optical lithography and Ar ion milling. Ta(5)/Au(100) are grown by sputtering for electrical contact.

*Thin film characterization*

Magnetic properties of unpatterned films are measured by a QuantumDesign SQUID-VSM magnetometer with a maximum field of 70 kOe or a Lakeshore VSM magnetometer with a maximum field of 20 kOe. X-ray diffraction spectra are measured in a Rigaku SmartLab diffractometer with a Cu K$\alpha$ source in parallel beam configuration and with a graphite monochromator on the detector side.

*Device characterization*

In current-induced switching experiments, a dc current is swept using a Keithley 2400 source meter and the Hall voltage is measured with a 2182A nanovoltmeter. The measurement clock is 0.6 s. The hard-axis anisotropy field of the MnGa with PMA is estimated from the curvature of $R_{\text{xy}}$ as a function of an in-plane applied field The magneto-transport is characterized in a QuantumDesign Physical Properties Measurement System (PPMS) on a horizontal rotator using the Resistivity option. The temperature is varied from 2 K to 400 K with an applied field up to 140 kOe. For harmonic Hall measurements, a sinusoidal signal of constant amplitude and at a frequency of 512.32 Hz is applied using a DS360 low distortion function generator. The first and second harmonic Hall voltages are measured with two SR830 lock-in amplifiers. The root-mean-square current of the circuit is estimated by measuring the voltage drop across a 100 $\Omega$ series resistor using a third lock-in amplifier.

*First principles calculations*

We considered a B2 structure of stoichiometric $\beta$-CoGa, as shown in the inset of Figure 1(b), and assumed the experimentally determined lattice constant of 0.287 nm. First principles calculations were carried out based on generalized gradient approximation[38] by using the full-potential linearized augmented plane wave method[39], where a plane wave cut-off of 3.9 Bohr$^{-1}$ has been used and muffin tin radius of 2.2 Bohr for Co and Ga atoms have been chosen. The intrinsic spin Hall conductivity was evaluated by means of the Kubo-formula[40,41] in the static limit ($\omega = 0$), where 97,336 special k-points were used to suppress numerical errors.

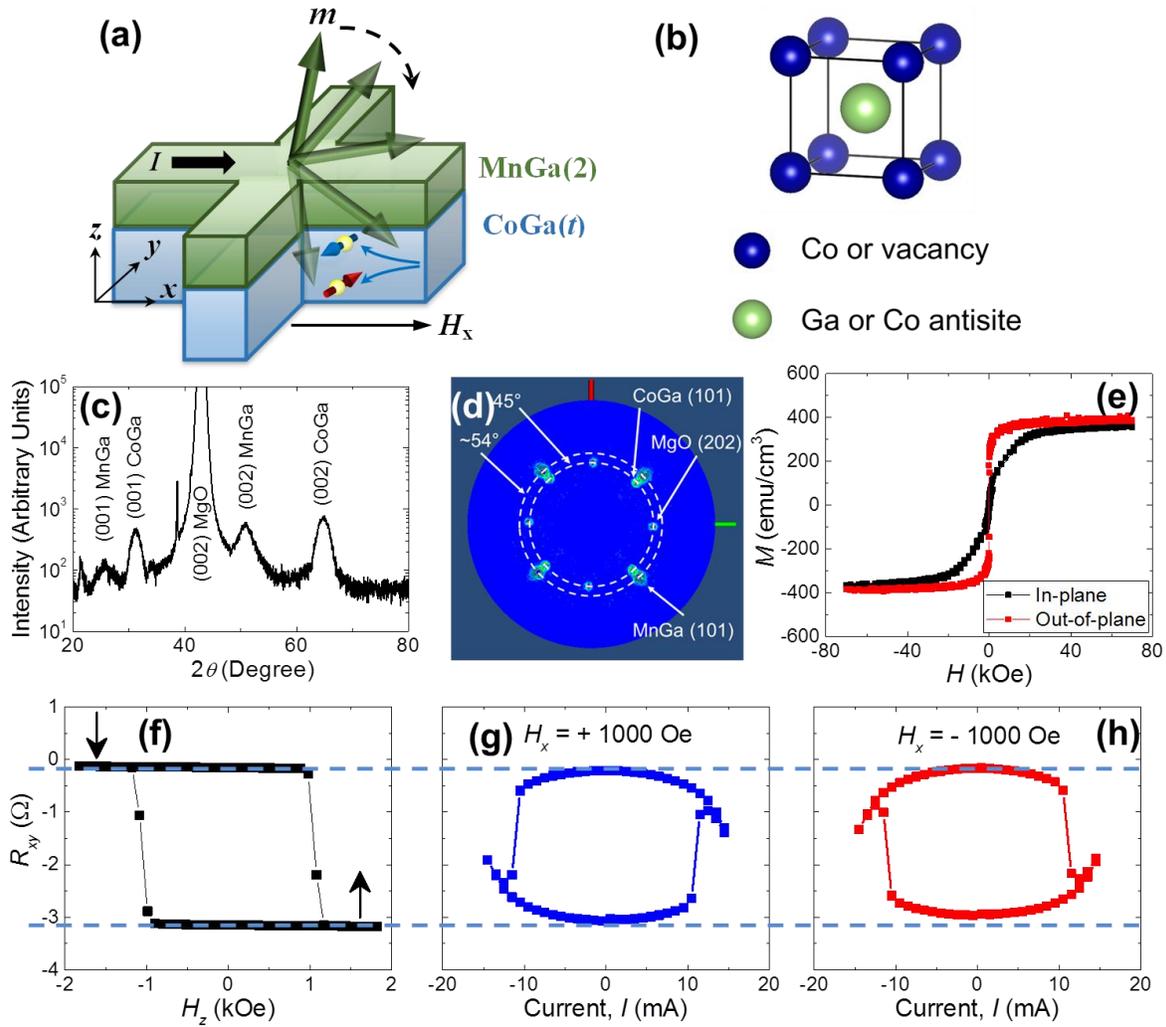

**Figure 1: Structure, magnetic properties and spin-orbit-torque-induced switching.**

(a) Schematic of the CoGa/FM bilayer structure with the definition of the coordinate system (*x*, *y*, *z*). (b) Illustration of the B2 $\beta$-CoGa crystal structure. (c) $\theta$-$2\theta$ X-ray diffraction spectrum of a CoGa(5)/Mn$_{1.44}$Ga(10) bilayer. (d) Pole figure of the same stack, revealing the in-plane epitaxial relationship. (e) In-plane and perpendicular *M-H* loops of a CoGa(5)/Mn$_{1.44}$Ga(2) heterostructure. (f) Typical $R_{xy}$-$H_z$ loop of a Hall bar device, confirming the strong PMA of MnGa. (g) and (h) Current-induced SOT switching of the MnGa magnetization, monitored via $R_{xy}$, with an applied *x*-field of ± 1 kOe. The two blue horizontal dashed lines indicate that practically all MnGa moments are reversed by the current.

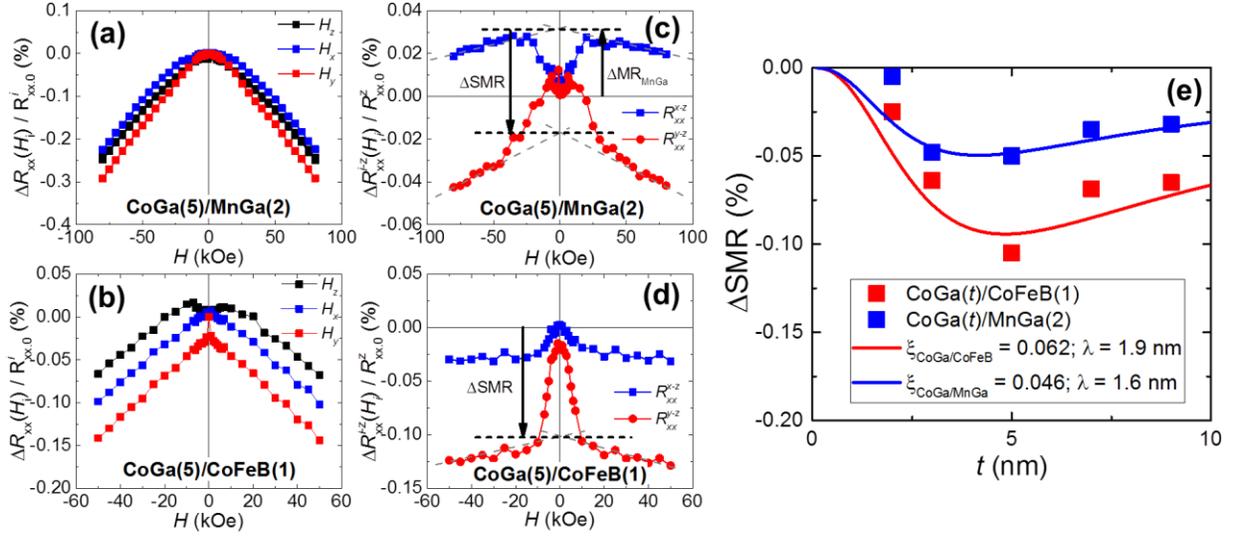

**Figure 2: Spin Hall magnetoresistance of CoGa/MnGa and CoGa/CoFeB bilayers.**

Magnetoresistance as a function of three orthogonal applied fields for (a) CoGa(5)/MnGa(2) and (b) CoGa(5)/CoFeB(1) heterostructures. (c) and (d) $\Delta R_{xx}^{x-z}/R_{xx,0}^z$ and $\Delta R_{xx}^{y-z}/R_{xx,0}^z$, calculated from (a) and (b), respectively. Different magnetoresistance contributions are labeled. (e) CoGa thickness ($t$) dependence of the SMR ratio, $\Delta$SMR. Solid lines are fits to the data using Equation (1).

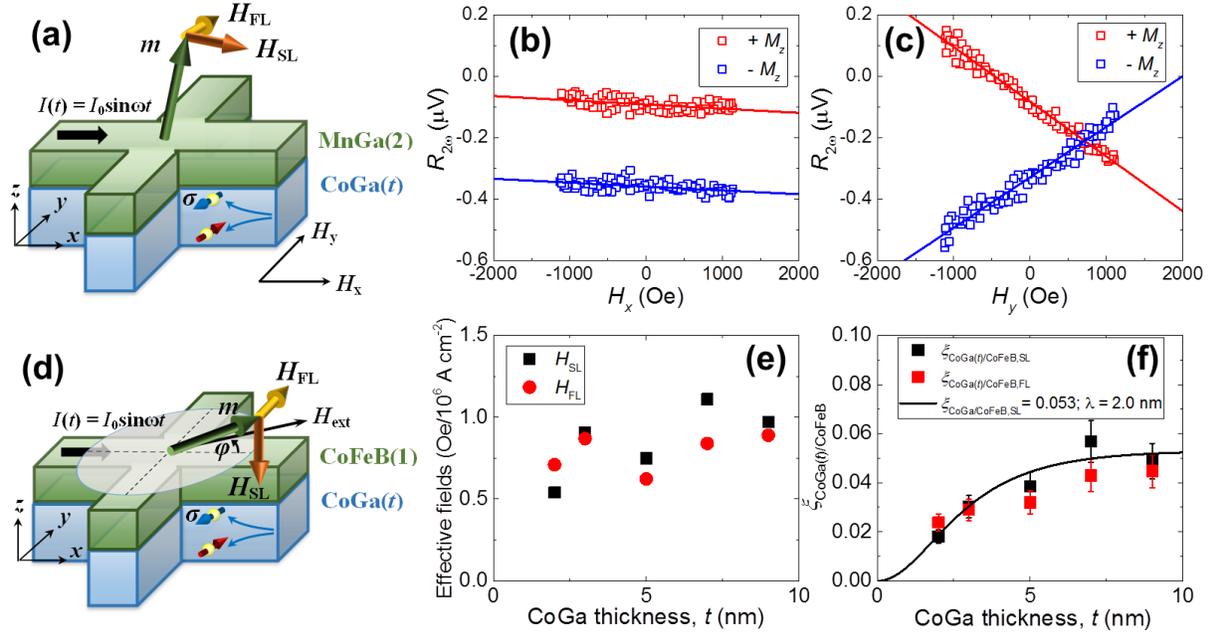

**Figure 3: Harmonic Hall measurements for quantifying spin-orbit effective fields.**

(a) Schematic of the circuit for harmonic Hall measurements on CoGa(5)/MnGa(2) bilayer with PMA. Brown and yellow arrows indicate the Slonczewski-like ($H_{SL}$) and the field-like ($H_{FL}$) spin-orbit effective field vector, respectively, due to an applied charge current flowing along $+x$. (b) and (c) $V_{2\omega}$ against an applied field along $x$ ($H_x$) and along $y$ ($H_y$). (d) Schematic of the circuit for harmonic Hall measurements on in-plane magnetized CoGa($t$)/CoFeB(1) bilayers. (e) CoGa thickness dependence of the spin-orbit effective fields. (f) Slonczewski-like $\xi_{SL}$ and field-like $\xi_{FL}$ spin Hall efficiencies versus $t$. Solid line is a fit on $\xi_{SL}$ using Equation (3).

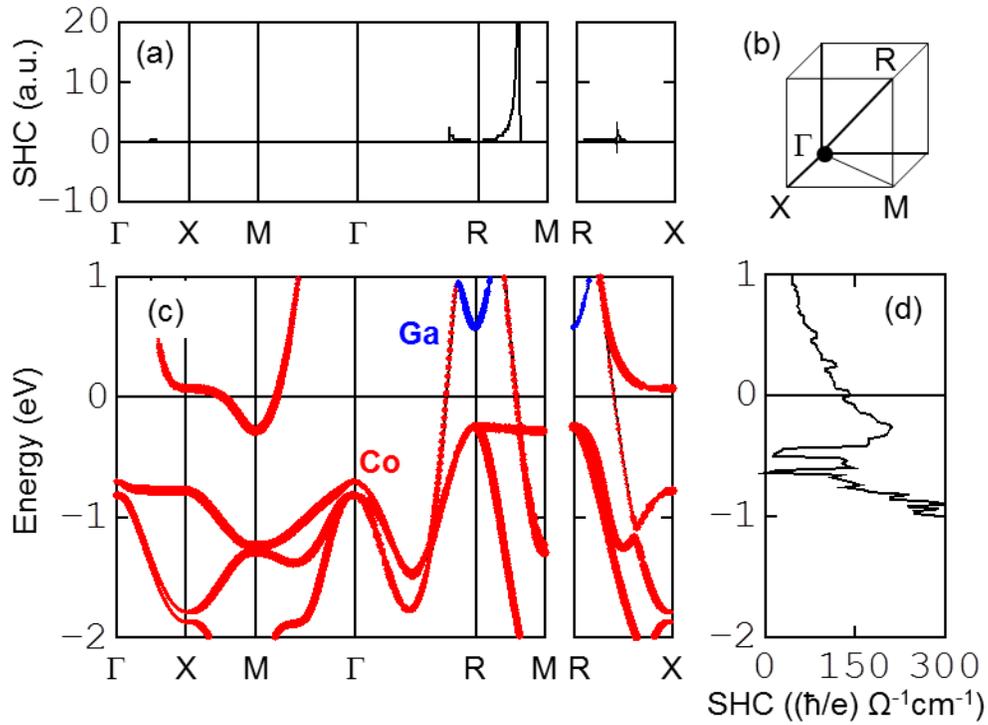

**Figure 4: Band structure and spin Hall conductivity of CoGa from first-principles calculations.**

(a) Contributions to the spin Hall conductivity, SHC, (in arbitrary unit) at different k-points. (b) Illustration of the first Brillouin zone of a simple cubic structure and the positions of high symmetry k-points. (c) Band structure along high symmetry directions where the reference energy ($E$=0) sets the Fermi level. (d) Calculated SHC against energy within a rigid band model.

# References

1       Miron, I. M. *et al.* Perpendicular switching of a single ferromagnetic layer induced by in-plane current injection. *Nature* **476**, 189-193 (2011).

2       Liu, L. *et al.* Spin-Torque Switching with the Giant Spin Hall Effect of Tantalum. *Science* **336**, 555-558 (2012).

3       Fan, Y. *et al.* Magnetization switching through giant spin–orbit torque in a magnetically doped topological insulator heterostructure. *Nat Mater* **13**, 699-704 (2014).

4       Lau, Y.-C., Betto, D., Rode, K., Coey, J. M. D. & Stamenov, P. Spin–orbit torque switching without an external field using interlayer exchange coupling. *Nat Nano* **11**, 758-762 (2016).

5       Fukami, S., Zhang, C., DuttaGupta, S., Kurenkov, A. & Ohno, H. Magnetization switching by spin-orbit torque in an antiferromagnet-ferromagnet bilayer system. *Nat Mater* **15**, 535-541 (2016).

6       Oh, Y.-W. *et al.* Field-free switching of perpendicular magnetization through spin–orbit torque in antiferromagnet/ferromagnet/oxide structures. *Nat Nano* **11**, 878-884 (2016).

7       Niimi, Y. *et al.* Giant Spin Hall Effect Induced by Skew Scattering from Bismuth Impurities inside Thin Film CuBi Alloys. *Physical Review Letters* **109**, 156602 (2012).

8       Sánchez, J. C. R. *et al.* Spin-to-charge conversion using Rashba coupling at the interface between non-magnetic materials. *Nature Communications* **4**, 2944 (2013).

9       Mellnik, A. R. *et al.* Spin-transfer torque generated by a topological insulator. *Nature* **511**, 449-451 (2014).

10     Pai, C. F. *et al.* Spin transfer torque devices utilizing the giant spin Hall effect of tungsten. *Applied Physics Letters* **101**, 122404 (2012).

11     Niimi, Y. *et al.* Giant Spin Hall Effect Induced by Skew Scattering from Bismuth Impurities inside Thin Film CuBi Alloys. *Physical Review Letters* **109**, 156602 (2012).

12     Hoffmann, A. Spin Hall Effects in Metals. *Ieee Transactions on Magnetics* **49**, 5172-5193 (2013).

13     Liu, J., Ohkubo, T., Mitani, S., Hono, K. & Hayashi, M. Correlation between the spin Hall angle and the structural phases of early 5d transition metals. *Applied Physics Letters* **107**, 232408 (2015).

14     Guo, G. Y., Murakami, S., Chen, T. W. & Nagaosa, N. Intrinsic Spin Hall Effect in Platinum: First-Principles Calculations. *Physical Review Letters* **100**, 096401 (2008).

15     Tanaka, T. *et al.* Intrinsic spin Hall effect and orbital Hall effect in 4d and 5d transition metals. *Physical Review B* **77**, 165117 (2008).

16     Graf, T., Felser, C. & Parkin, S. S. P. Simple rules for the understanding of Heusler compounds. *Progress in Solid State Chemistry* **39**, 1-50 (2011).

17     Balke, B., Fecher, G. H., Winterlik, J. & Felser, C. Mn3Ga, a compensated ferrimagnet with high Curie temperature and low magnetic moment for spin torque transfer applications. *Applied Physics Letters* **90**, 152504 (2007).

18     Suzuki, K. Z., Ranjbar, R., Sugihara, A., Miyazaki, T. & Mizukami, S. Room temperature growth of ultrathin ordered MnGa films on a CoGa buffer layer. *Japanese Journal of Applied Physics* **55**, 010305 (2016).

19     Mizukami, S. *et al.* Long-Lived Ultrafast Spin Precession in Manganese Alloys Films with a Large Perpendicular Magnetic Anisotropy. *Physical Review Letters* **106**, 117201 (2011).

20     Kurt, H., Rode, K., Venkatesan, M., Stamenov, P. & Coey, J. M. D. High spin polarization in epitaxial films of ferrimagnetic Mn${}_{3}$Ga. *Physical Review B* **83**, 020405 (2011).

21     Jeong, J. *et al.* Termination layer compensated tunnelling magnetoresistance in ferrimagnetic Heusler compounds with high perpendicular magnetic anisotropy. *Nat Commun* **7**, 10276 (2016).

22    Berner, D., Geibel, G., Gerold, V. & Wachtel, E. Structural defects and magnetic properties in the ordered compound CoGa. *Journal of Physics and Chemistry of Solids* **36**, 221-227 (1975).

23    Zhang, W., Han, W., Jiang, X., Yang, S.-H. & S. P. Parkin, S. Role of transparency of platinum-ferromagnet interfaces in determining the intrinsic magnitude of the spin Hall effect. *Nat Phys* **11**, 496-502 (2015).

24    Nakayama, H. *et al.* Spin Hall Magnetoresistance Induced by a Nonequilibrium Proximity Effect. *Physical Review Letters* **110**, 206601 (2013).

25    Althammer, M. *et al.* Quantitative study of the spin Hall magnetoresistance in ferromagnetic insulator/normal metal hybrids. *Physical Review B* **87**, 224401 (2013).

26    Kim, J., Sheng, P., Takahashi, S., Mitani, S. & Hayashi, M. Spin Hall Magnetoresistance in Metallic Bilayers. *Physical Review Letters* **116**, 097201 (2016).

27    Chen, Y.-T. *et al.* Theory of spin Hall magnetoresistance. *Physical Review B* **87**, 144411 (2013).

28    Kim, J. *et al.* Layer thickness dependence of the current-induced effective field vector in Ta│CoFeB│MgO. *Nat Mater* **12**, 240-245 (2013).

29    Garello, K. *et al.* Symmetry and magnitude of spin-orbit torques in ferromagnetic heterostructures. *Nat Nano* **8**, 587-593 (2013).

30    Avci, C. O. *et al.* Interplay of spin-orbit torque and thermoelectric effects in ferromagnet/normal-metal bilayers. *Physical Review B* **90**, 224427 (2014).

31    Hayashi, M., Kim, J., Yamanouchi, M. & Ohno, H. Quantitative characterization of the spin-orbit torque using harmonic Hall voltage measurements. *Physical Review B* **89**, 144425 (2014).

32    Liu, L., Moriyama, T., Ralph, D. C. & Buhrman, R. A. Spin-Torque Ferromagnetic Resonance Induced by the Spin Hall Effect. *Physical Review Letters* **106**, 036601 (2011).

33    Manchon, A., Koo, H. C., Nitta, J., Frolov, S. M. & Duine, R. A. New perspectives for Rashba spin-orbit coupling. *Nature Materials* **14**, 871-882 (2015).

34    Saidaoui, H. B. & Manchon, A. Spin-Swapping Transport and Torques in Ultrathin Magnetic Bilayers. *Physical Review Letters* **117**, 036601 (2016).

35    Nakatsuji, S., Kiyohara, N. & Higo, T. Large anomalous Hall effect in a non-collinear antiferromagnet at room temperature. *Nature* **527**, 212-215 (2015).

36    Nayak, A. K. *et al.* Large anomalous Hall effect driven by a nonvanishing Berry curvature in the noncolinear antiferromagnet Mn3Ge. *Science Advances* **2**, e1501870 (2016).

37    Zhang, Y. *et al.* Strong anisotropic anomalous Hall effect and spin Hall effect in the chiral antiferromagnetic compounds Mn3X (X = Ge, Sn, Ga, Ir, Rh, and Pt). *Physical Review B* **95**, 075128 (2017).

38    Perdew, J. P., Burke, K. & Ernzerhof, M. Generalized Gradient Approximation Made Simple. *Physical Review Letters* **77**, 3865-3868 (1996).

39    Nakamura, K., Ito, T., Freeman, A. J., Zhong, L. & Fernandez-de-Castro, J. Enhancement of magnetocrystalline anisotropy in ferromagnetic Fe films by intra-atomic noncollinear magnetism. *Physical Review B* **67**, 014420 (2003).

40    Oppeneer, P. M., Maurer, T., Sticht, J. & Kübler, J. Ab initio. *Physical Review B* **45**, 10924-10933 (1992).

41    Guo, G. Y., Yao, Y. & Niu, Q. Ab initio Calculation of the Intrinsic Spin Hall Effect in Semiconductors. *Physical Review Letters* **94**, 226601 (2005).

**Supplementary information of**

**Spin Hall effect from hybridized 3*d*-4*p* orbitals**

Yong-Chang Lau[1,2†**], Hwachol Lee[1†], Kohji Nakamura[3] and Masamitsu Hayashi[1,2**]

[1]*National Institute for Materials Science, Tsukuba 305-0047, Japan*

[2]*Department of Physics, The University of Tokyo, Bunkyo, Tokyo 113-0033, Japan*

[3]*Department of Physics Engineering, Mie University, Tsu, Mie 514-8507, Japan*

[†]These authors contributed equally to this work

[*]yongchang.lau@qspin.phys.s.u-tokyo.ac.jp

[**]hayashi@phys.s.u-tokyo.ac.jp

**S1. Ordering parameter of B2 *β*-CoGa**

**S2. Magnetoresistance in CoGa(5)/MAO(2)/Ta(1) heterostructure**

**S3. Magnetoresistance in MnGa(20)/MAO(2)/Ta(1) heterostructure**

**S4. Extraction of SMR ratio, ΔSMR in CoGa(*t*)/FM bilayer systems**

**S5. Harmonic Hall measurement for in-plane magnetized CoGa(*t*)/CoFeB**

**S6. Estimation of the electron mean free path**

## S1. Ordering parameter of *B*2 *β*-CoGa

In *B*2 structure, the ordering is given by the differentiation of atomic occupancy on 1*a* (0, 0, 0) and 1(b) (0.5, 0.5, 0.5) sites (Figure 1(b) of the manuscript). The ordering can be revealed by comparing the X-ray diffraction intensity ratio of a superlattice peak (e.g. (001) reflection) and a fundamental one (e.g. (002) reflection). In perfectly ordered equiatomic $Co_{50}Ga_{50}$ compound, a moderate (001)/(002) intensity ratio of ~0.2 is expected, due to the small difference in atomic number $Z$ between Co and Ga ($Z$(Co) ~ 59 and $Z$(Ga) ~ 69.7). In a fully disordered bcc crystal, with no difference in atomic occupancy on 1*a* and 1*b* sites, the (001) reflection is forbidden and the (001)/(002) intensity ratio is 0. The ordering parameter $S$ is given by the following expression:

$$S = \sqrt{\frac{(001)/(002)\text{ratio}_{\text{exp}}}{(001)/(002)\text{ratio}_{\text{cal}}}} \qquad (1)$$

If the vacancy formation at 1*a* is allowed, it can be regarded as a third species with $Z$(vacancy) = 0 and with *unknown* concentration. It largely increases the atomic contrast between the two sites and can give rise to an intensity ratio > 0.2. However, for a given experimental (001)/(002) ratio, there will be no unique solution of the corresponding atomic ordering, without knowing the concentration of the vacancy.

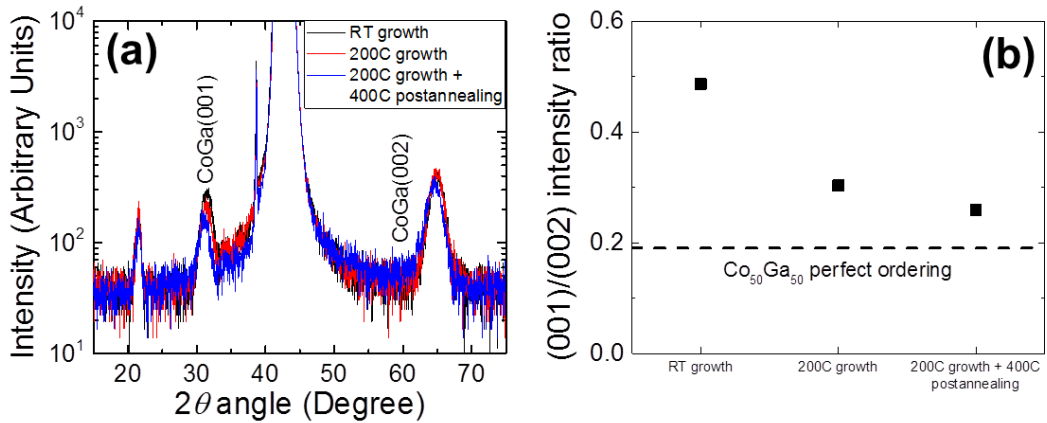

Figure S1: (a) X-ray diffraction spectra of CoGa(5)/CoFeB(1) bilayers with CoGa grown under different conditions. (b) Comparison between the experimental (001)/(002) intensity ratio and that expected from a fully ordered CoGa.

Figure S1(a) shows the symmetrical $\theta$-$2\theta$ X-ray diffraction spectra of three CoGa(5)/CoFeB(1) samples with different CoGa growth conditions: Growth at room temperature (RT), growth at $200^o$C, and the optimized growth at $200^o$C followed by postannealing at $400^o$C for 30 minutes. The three samples exhibit similar (002) peak intensity which indicates that their crystallinity is comparable, whereas the evolution of (001) peak intensity reveals the change of ordering parameter. The experimental (001)/(002) peak intensity ratios are plotted in Figure S1(b). The data are all higher than the calculated intensity ratio of perfect $Co_{50}Ga_{50}$ (dashed horizontal line), which suggests in all samples the presence of vacancy. The density of vacancy in CoGa crystallites decreases upon introducing thermal treatment. We estimate that the optimized $Co_{53}Ga_{47}$ films used in this work may contain up to 10% vacancy on the $1a$ site and with ~10 % of $1b$ being occupied by Co, in lieu of Ga.

## S2. Magnetoresistance in CoGa(5)/MAO(2)/Ta(1) heterostructure

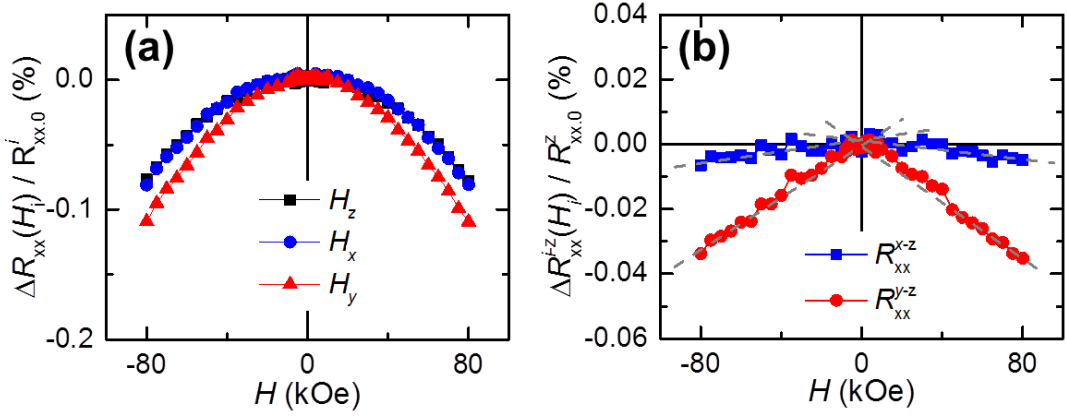

Figure S2: $\Delta R_{xx}(H_i)/R_{xx,0}^i$ (a) and $\Delta R_{xx}^{i-z}(H_i)/R_{xx,0}^z$ (b) as a function of applied field $H_i$ along $x$, $y$ and $z$ directions at $T = 295$ K for the CoGa(5) layer.

We measure the longitudinal resistance $R_{xx}$ of MgO(001) substrate/CoGa(5)/MAO(2)/Ta(1) stack as a function of applied field $H_i$ along three orthogonal directions: $i = x$ (in-plane parallel to current), $i = y$ (in-plane transverse to current), and $i = z$ (out-of-plane transverse to current). We use the same notation as in the main text, $\Delta R_{xx}(H_i) \equiv R_{xx}(H_i) - R_{xx,0}^i$ and $\Delta R_{xx}^{i-z}(H_i) \equiv R_{xx}(H_i) - R_{xx}(H_z)$ where $R_{xx,0}^i \equiv R_{xx}(H_i = 0)$ is the base resistance. $\Delta R_{xx}(H_i)/R_{xx,0}^i$ and $\Delta R_{xx}^{i-z}(H_i)/R_{xx,0}^z$ as a function of applied field $H_i$ at $T = 295$ K are plotted in Figure S2(a) and (b), respectively. The CoGa single layer exhibits negative magnetoresistance (MR) in all directions with quadratic-like field dependence. In addition, there is a small directional MR anisotropy of the form $R_{xx}(H_x) \sim R_{xx}(H_z) > R_{xx}(H_y)$ that does not saturate at 80 kOe. From Figure S2(b), $\Delta R_{xx}^{y-z}(H)$ is essentially linear with $H$ whereas $\Delta R_{xx}^{x-z} \approx 0$. Although we do not know the origin of this MR contribution, we argue that it may contain a small contribution from Hanle magnetoresistance (HMR)[1], which has the same field directional dependence but scales with $H^2$. The minor contribution from HMR is also understood from the magnitude of the observed MR ($\sim 4 \times 10^{-4}$ @ 80 kOe), being significantly higher than the reported HMR values in YIG/Pt(7) ($\sim 6 \times 10^{-5}$ @ 90 kOe) or SiO$_2$/Ta(5) ($\sim 4.5 \times 10^{-6}$ @ 90 kOe) structures.

We perform linear extrapolation of $\Delta R_{xx}^{y-z}(H)$ and $\Delta R_{xx}^{x-z}(H)$ from high fields to zero field as shown in Figure S2(b). The two $y$-intercepts $R_{xx,0}^{x-z}$ and $R_{xx,0}^{y-z}$ practically coincide with the origin of the graph. We will apply this protocol hereafter on CoGa/FM bilayer systems in order to eliminate the field-dependent MR contribution from CoGa.

## S3. Magnetoresistance in MnGa(20)/MAO(2)/Ta(1) heterostructure

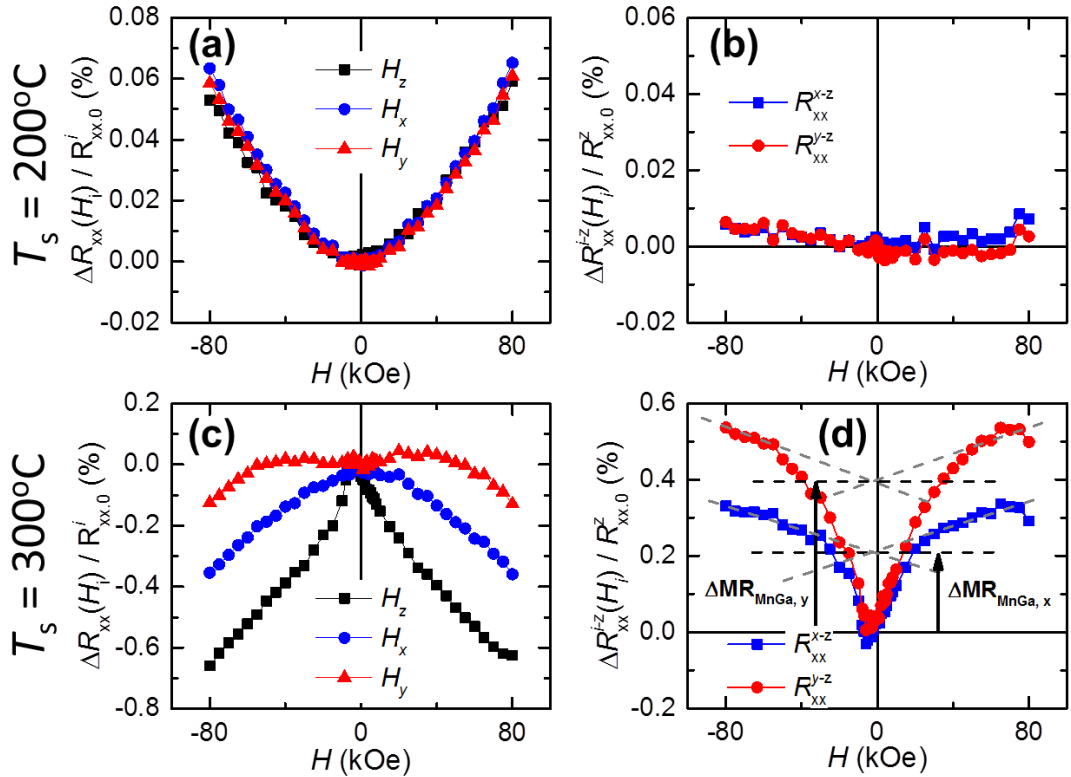

Figure S3: $\Delta R_{xx}(H_i)/R_{xx,0}^i$ (a&c) and $\Delta R_{xx}^{i-z}(H_i)/R_{xx,0}^z$ (b&d) as a function of applied field $H_i$ along $x$, $y$ and $z$ directions. Results for MnGa(20) films being deposited at a substrate

temperature of $T_s = 200^oC$ (a&b) and $T_s = 300^oC$ (c&d) are compared. Note that the scales of the upper and lower panels differ by a factor of 10.

It is not trivial to extract the magneto-transport contribution of MnGa grown on CoGa seed layer because of two reasons. First, the magnetoresistance (MR) of tetragonal MnGa with perpendicular magnetic anisotropy (PMA) depends sensitively on the atomic ordering of the compound. Second, to date, no insulating seed layer or substrate is known to promote the growth of MnGa of comparable quality with that grown on CoGa seed layer, under similar growth conditions. We grow thick (thickness ~ 20 nm) MnGa films directly on MgO substrates at elevated temperature, in an attempt to mimic the MR behavior of high-quality 2 nm MnGa grown on CoGa seed layer at ambient temperature. At $T_s = 200^oC$, the resulting MnGa film is practically non-magnetic. At $T_s = 300^oC$, the MnGa film starts to crystallize and exhibits PMA. Further increase of the substrate temperature improves the MnGa crystallinity but degrades the wetting and eventually leads to island-like insulating films. Based on the resistivity and the magnetic properties, we consider that the MnGa film grown at $T_s = 300^oC$ is the closest, in terms of transport properties, to the MnGa film grown on CoGa seed layer.

We perform similar field-dependent MR measurement on MgO(001) substrate/MnGa(20)/MAO(2)/Ta(1) heterostructures. Results are plotted in Figure S3(a)&(b) and Figure S4(c)&(d) for samples prepared at $T_s = 200^oC$ and $300^oC$, respectively. MnGa grown at $T_s = 200^oC$ exhibits weak positive MR with a quadratic field dependence and a very small field directional anisotropy < 0.01 %, as shown in Figure S3(b). In contrast, MnGa grown at $T_s = 300^oC$ shows negative MR that strongly depends on the field directions. The MR is the form $R_{xx}(H_y) > R_{xx}(H_x) > R_{xx}(H_z)$. We define $\Delta MR_{MnGa,i}$ as the zero-field MR contribution of MnGa upon extrapolating $\Delta R_{xx}^{i-z}(H)$ from high field. We obtain $\Delta MR_{MnGa,y} \approx 2$ x $\Delta MR_{MnGa,x} >$ $\Delta MR_{MnGa,x}$. A possible source of $\Delta MR_{MnGa}$ is the tetragonal lattice of MnGa which naturally leads to transport anisotropy between in-plane ($H_x$ and $H_y$) and out-of-plane ($H_z$) field directions. For simplicity, we will assume hereafter $\Delta MR_{MnGa,y} \approx \Delta MR_{MnGa,x} \equiv \Delta MR_{MnGa}$ in CoGa/MnGa bilayers. We will discuss the consequences of such assumption in Section S4.

## S4. Extraction of SMR ratio, $\Delta SMR$ in CoGa($t$)/FM bilayer systems

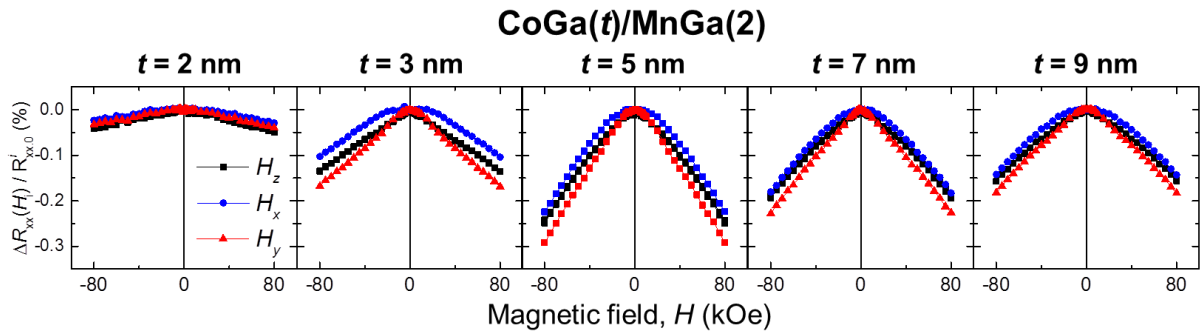

Figure S4: Field dependence of $\Delta R_{xx}(H_i)/R_{xx,0}^i$ for CoGa($t$)/MnGa(2) heterostructures with $t$ varying from 2 nm to 9 nm.

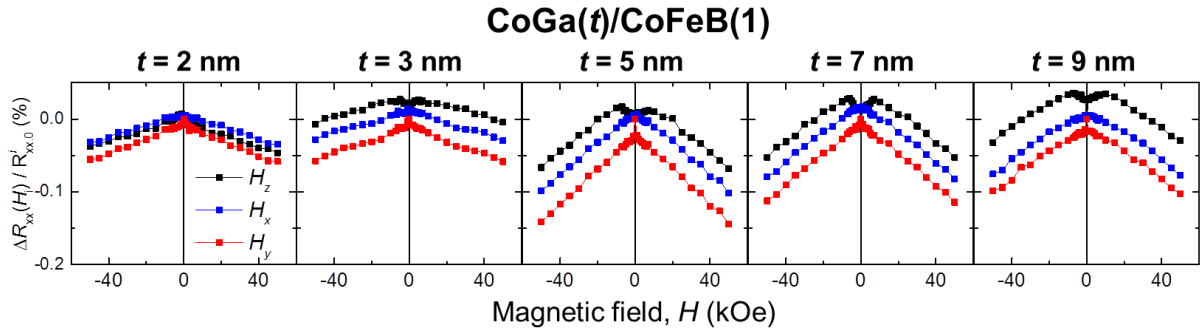

Figure S5: Field dependence of $\Delta R_{xx}(H_i)/R_{xx,0}^i$ for CoGa($t$)/CoFeB(1) heterostructures with $t$ varying from 2 nm to 9 nm.

The field dependence of $\Delta R_{xx}(H_i)/R_{xx,0}^i$ for CoGa($t$)/MnGa(2) and CoGa($t$)/CoFeB(1) heterostructures are plotted in Figure S4 and S5 respectively. For the CoGa($t$)/CoFeB(1) series, the non-saturating MR at high field is mainly arising from that of the CoGa layer (Section S2). For the CoGa($t$)/MnGa(2) series, the stronger high field MR agrees with a situation involving the additive MR contribution from the CoGa and the MnGa layers.

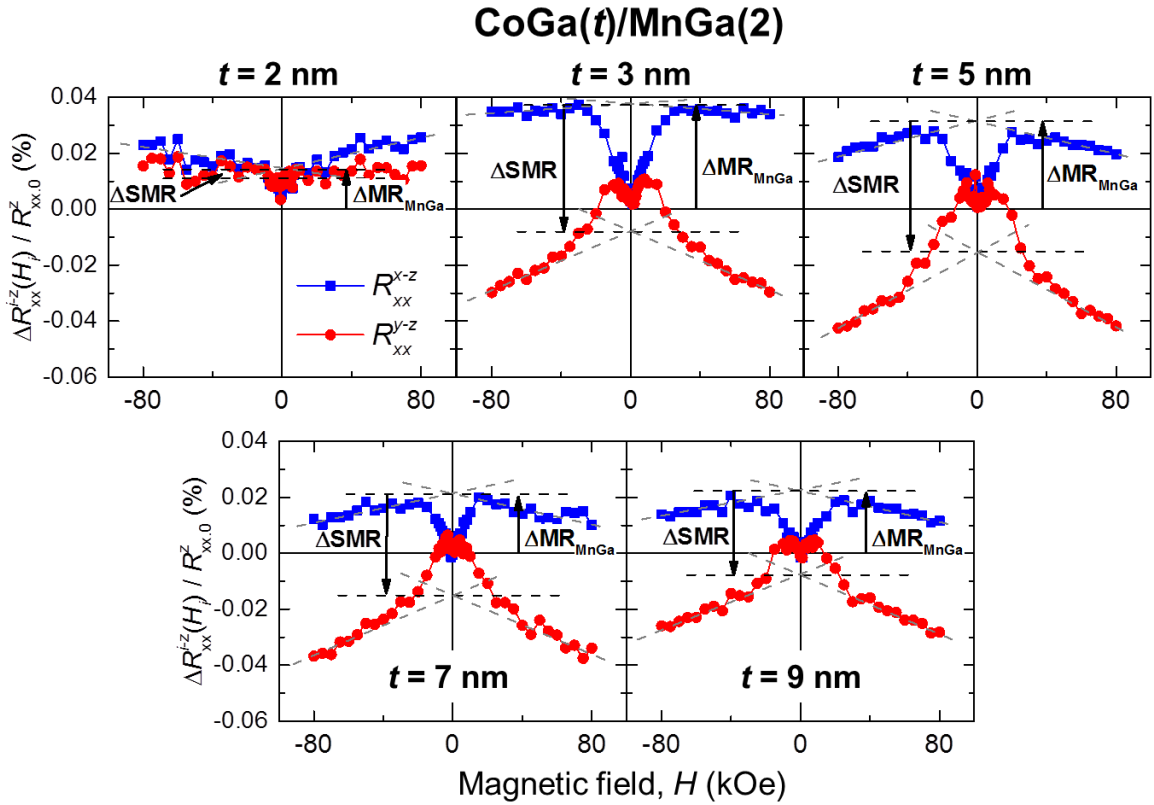

Figure S6: $t$ dependence of $\Delta R_{xx}^{i-z}(H_i)/R_{xx,0}^z$ as a function of applied field $H_i$ for CoGa($t$)/MnGa(2) samples. Contributions of $\Delta MR_{MnGa}$ and $\Delta SMR$ are shown for each thickness.

This high-field quasi isotropic MR can be largely eliminated by plotting $\Delta R_{xx}^{i-z}(H_i)/R_{xx,0}^z$ as a function of applied field $H_i$ for CoGa($t$)/MnGa(2) samples, as shown in Figure S6. We are interested in the MR contributions that depend on the magnetization direction of the adjacent

MnGa, which can be extracted by extrapolating the data from high field to zero field. We define $R_{xx,0}^{i-z}$ as the resulting $y$-intercept from the linear extrapolation. Following Section S3, we consider:

$$R_{xx,0}^{x-z} = \Delta MR_{MnGa} \quad (2)$$

$$R_{xx,0}^{y-z} \approx \Delta MR_{MnGa} + \Delta SMR \quad (3)$$

Since $\Delta MR_{MnGa,y}$ is positive and is likely to be larger than $\Delta MR_{MnGa,x}$, we tend to underestimate the size of $\Delta SMR$, which is negative.

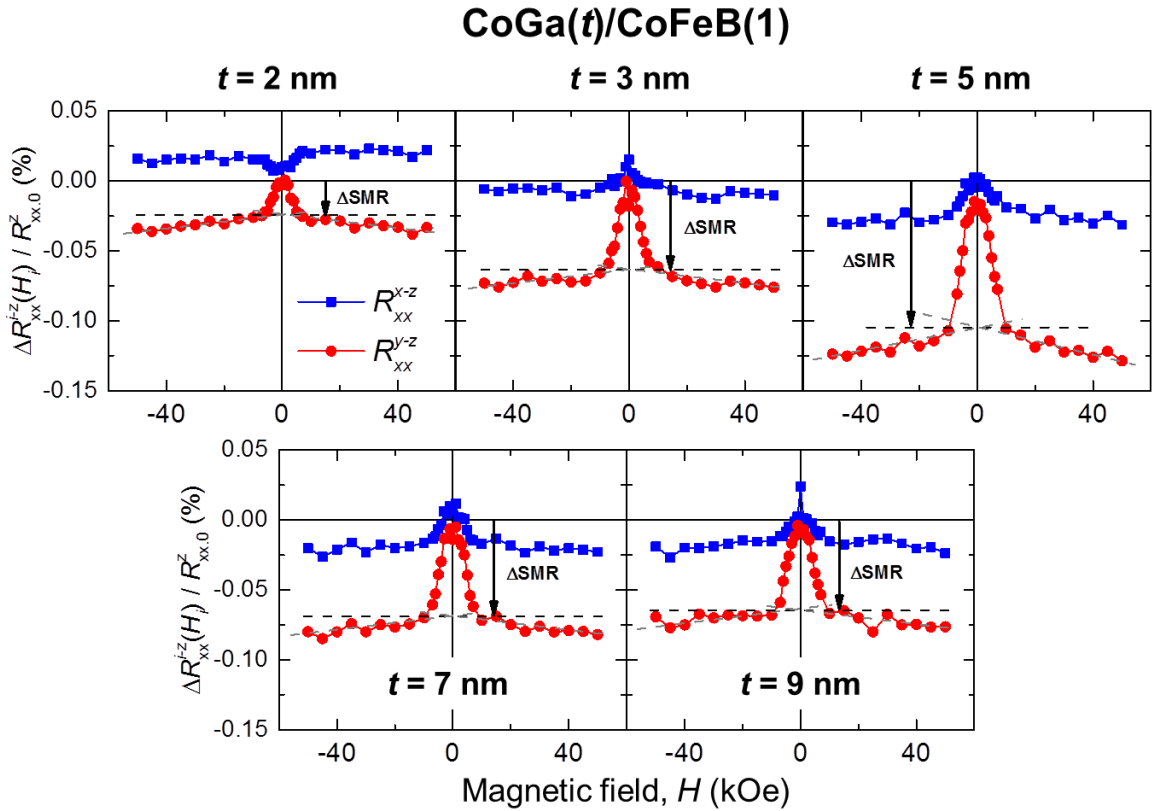

Figure S7: $t$ dependence of $\Delta R_{xx}^{i-z}(H_i)/R_{xx,0}^z$ as a function of applied field $H_i$ for CoGa($t$)/CoFeB(1) samples. The $\Delta SMR$ contribution is shown for each CoGa thickness.

$\Delta R_{xx}^{i-z}(H_i)/R_{xx,0}^z$ as a function of applied field $H_i$ for CoGa($t$)/CoFeB(1) samples are plotted in Figure S7. For simplicity, we consider:

$$R_{xx,0}^{x-z} \approx \Delta AMR \quad (4)$$

$$R_{xx,0}^{y-z} \approx \Delta SMR \quad (5)$$

$\Delta SMR$ for each CoGa thickness are shown in Figure S7.

We should also note that for $t \geq 3$ nm, the negative sign of $R_{xx,0}^{x-z}$ is inconsistent with the positive AMR of CoFeB[2]. The origin of this unconventional negative AMR is unclear for the moment. We speculate that the unintentional intermixing at the CoGa/CoFeB interface during the growth may have formed an alloy with different magneto-transport properties, compared to CoGa and CoFeB alone.

## S5. Harmonic Hall measurement for in-plane magnetized CoGa($t$)/CoFeB

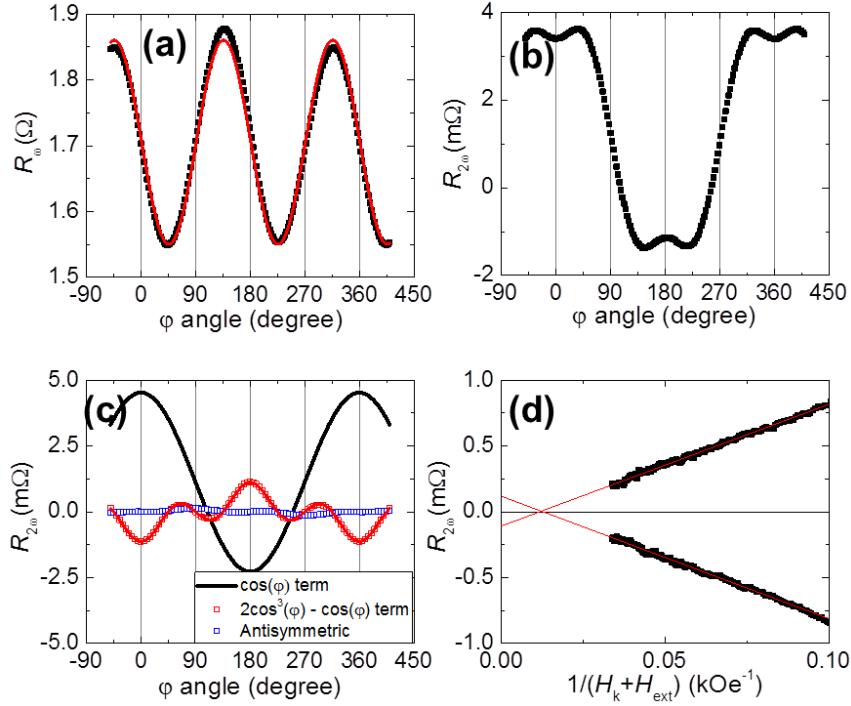

Figure S8: (a) $R_\omega$ as a function of external field direction $\varphi$. Red line is a sin2$\varphi$ fit to the data. (b) Raw $R_{2\omega}$ as a function of external field direction $\varphi$. (c) Decomposition of $R_{2\omega}$ into a "cos $\varphi$" term and a "$2\cos^3\varphi - \cos\varphi$" term. (d) $R_{2\omega}$ against $\frac{1}{H_k + |H_{ext}|}$ for $H_{ext}$ along $\varphi = 45^o$ and $225^o$. Red lines are linear fits to the data.

We extract the current-induced spin-orbit effective fields from the angular dependence of $R_{2\omega}$, as described by Avci $et\ al.$[3] Considering the case where both the external field $\boldsymbol{H}_{ext}$ and the magnetization vector $\boldsymbol{m}$ lie in the ($xy$) plane, $i.e.$ the polar angle $\theta = 90^o$. The expressions of the harmonic Hall resistances are of the form:

$$R_\omega = R_{AHE}\cos\theta + R_{PHE}\sin^2\theta\sin 2\varphi \quad (6)$$

$$R_{2\omega} = -\left(R_{\text{AHE}} \frac{H_{\text{SL}}}{H_{\text{ext}}+H_k} + R_{\nabla T}^0\right)\cos\varphi + 2R_{\text{PHE}}(2\cos^3\varphi - \cos\varphi)\frac{H_{\text{FL}}+H_{\text{Oe}}}{H_{\text{ext}}} \equiv R_{\text{SL}} +$$

$$R_{\text{FL+Oe}} + R_{\nabla T} \qquad (7)$$

where $R_{\text{SL}}$, $R_{\text{FL+Oe}}$, and $R_{\nabla T}$ are Slonczewski-like effective field, field-like effective field + Oersted field, and thermoelectric contribution to $R_{2\omega}$, respectively. $R_{\text{AHE}}$, $R_{\text{PHE}}$ and $H_k$ denote anomalous Hall resistance, planar Hall resistance and the anisotropy field of the CoFeB layer. Typical $R_\omega(\varphi)$ and $R_{2\omega}(\varphi)$ for a CoGa(5)/CoFeB(1)/MAO(2)/Ta(1) film measured with $H_{\text{ext}} =$ 1300 Oe are plotted in Figure S8 (a) and (b). The prefactor of the "$\cos\varphi$" component can be extracted by fitting $R_{2\omega}(\varphi)$ to a $\cos\varphi$ function at $\varphi = 45^o$, $135^o$, $225^o$ and $315^o$, where at these particular points the "$2\cos^3\varphi - \cos\varphi$" component vanishes. Then, upon subtracting the "$\cos\varphi$" contribution from $R_{2\omega}(\varphi)$, we fit the data with a $2\cos^3\varphi - \cos\varphi$ function. A decomposition of $R_{2\omega}(\varphi)$ into the two contributions are shown in Figure S8 (c). We note that the antisymmetric component of the raw $R_{2\omega}(\varphi)$ data, which mainly arises from the sample misalignment, is very low.

Based on Eq. 7, the prefactor of the "$\cos\varphi$" component may contain a thermoelectric contribution $R_{\nabla T}^0$ that does not depend on $H_{\text{ext}}$. We measure $R_{2\omega}(H_{\text{ext}})$ with the field applied along $\varphi = 45^o$ and $225^o$. We plot $R_{2\omega}$ against $\frac{1}{H_k+|H_{\text{ext}}|}$ in Figure S8 (d). We find a relatively small $R_{\nabla T}^0$ from the $y$-intercept, which will increase $H_{\text{SL}}$ by ~ 10 %.

# S6. Estimation of the electron mean free path

The electron mean free path $l$ in CoGa can be estimated using an expression derived from the free electron model for a three dimensional system[4]:

$$l = \frac{3\pi^2 \hbar}{e^2 k_F^2 \rho}$$

$\rho$ is the longitudinal resistivity of CoGa and $k_F = \sqrt[3]{3\pi^2 n}$ is the Fermi wave vector that depends on the effective carrier density, $n$. We assume an effective carrier density of ~1 electron per formula unit (or $n$ ~ 4.3 x $10^{28}$ carrier / $m^3$) for CoGa, which is typical for metals. Using $\rho = 175\ \mu\Omega$cm, we obtain $l$ ~ 0.59 nm, which is very short and is of the order of two unit cells ($d$ ~ 0.287 nm).

# References:

1    Vélez, S. *et al.* Hanle Magnetoresistance in Thin Metal Films with Strong Spin-Orbit Coupling. *Physical Review Letters* **116**, 016603 (2016).
2    Kim, J., Sheng, P., Takahashi, S., Mitani, S. & Hayashi, M. Spin Hall Magnetoresistance in Metallic Bilayers. *Physical Review Letters* **116**, 097201 (2016).
3    Avci, C. O. *et al.* Interplay of spin-orbit torque and thermoelectric effects in ferromagnet/normal-metal bilayers. *Physical Review B* **90**, 224427 (2014).
4    Gunnarsson, O., Calandra, M. & Han, J. E. Colloquium: Saturation of electrical resistivity. *Reviews of Modern Physics* **75**, 1085-1099 (2003).